# Spinodal Decomposition of Metastable Melting in the Zero-Temperature Limit

Igor L. Iosilevski, Alexander Yu. Chigvintsev

*Moscow Institute of Physics and Technology (State University) Dolgoprudny 141700, Russia*

## INTRODUCTION

Melting and freezing of Coulomb particles is common phenomena in many astrophysical objects [1,2,3]. One-component plasma model (OCP) in its standard variant with rigid compensating background (notified below as OCP(#)) is widely used for theoretical description of melting process of Coulomb (Wigner) crystal [4,5]. The modified OCP on uniform, but compressible background (notified below as OCP(~)) enriched noticeably combination of phase transitions in the model due to addition of 1[st]-order phase transitions of gas-liquid and gas-solid type with upper critical point [6]. Study of phase transitions in OCP(~) makes it possible to learn general picture of thermodynamically stable and metastable melting and freezing in wide range of thermodynamic parameters including in particular deeply metastable coexistence of extended crystal with overcooled liquid [7]. Recent progress in the technique of dynamic experiments of reaching deep negative pressure in metastable extended condensed matter in crystal and fluid states [8] raises a question about behavior of metastable melting in the limit of zero temperature. Important data to analyze this problem can be obtained from a numerical simulation of phase transitions in sufficiently realistic many-particles systems. Besides the analysis of the problem could be supplemented the study of idealized modeling systems where the features of phase transitions can be calculated directly due to simplified modeling nature of these systems. Present paper is devoted to study of various scenarios of hypothetical closure of zero-temperature limit melting. The analysis is based on the features of idealized models. First of all it is the line of so-called "non-associative" coulomb models [6,7,9] the number of modified versions of well-known prototype – one-component plasma model with rigid uniform compressible background.

### "Conventional" Scenarios of Metastable Melting in Ordinary Substances in the Limit $T \Rightarrow 0$

It is assumed according to this scenario (for example [10]) that metastable melting curve reaches the matter 'cold curve' (isotherm $T = 0$). It is assumed also that this hypothetical metastable branch of melting curve is in close agreement with so-called Simon's law (1). Phase diagram with such behavior is shown on the Figure 1.

$$P_{melting} = AT^C + P^*; \qquad (A, C \text{ and } P^* = \text{const.}) \qquad (1)$$

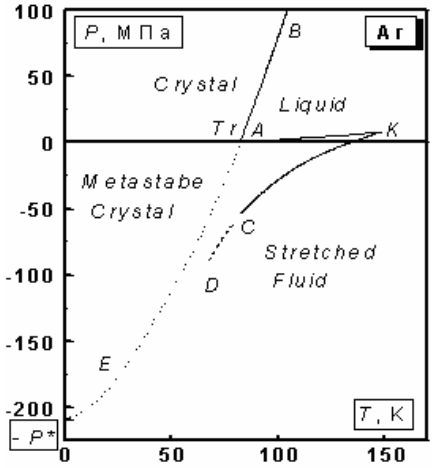 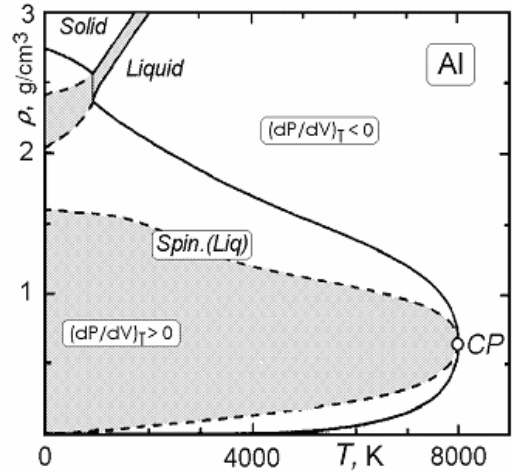

**FIGURE 1.** *P-T* phase diagram of argon with hypothetical metastable melting (figure after [10]): *AB* – melting curve; *AK* – boiling curve; *A* – triple point; *K* - critical point; *KCD* –fluid spinodal; *AEP\** – metastable branch of melting curve (Simon's law); *P\** – hypothetical zero-temperature limit of melting.

**FIGURE 2.** Density-temperature phase diagram of Al via semi-empirical EOS [11] constructed after recommendations [10] (Figure from [11]): *CP* - critical point; *M* – melting; *M\** – hypothetical metastable melting; *Spin(Liq)* – liquid spinodal $\{(\partial P/\partial V)_T = 0\}$; *SpG* – gas spinodal.

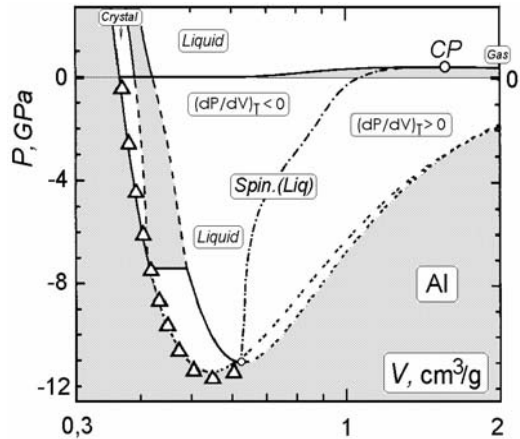 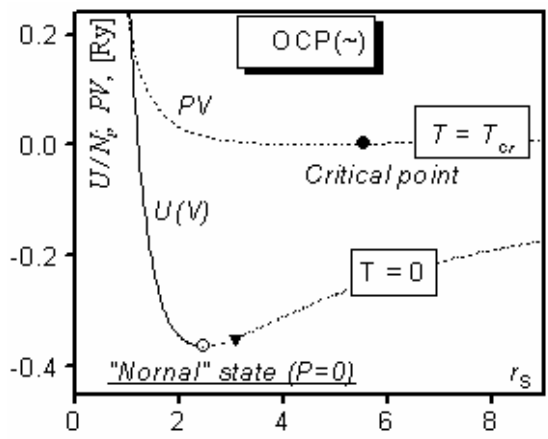

**FIGURE 3.** The same as Fig. 2 in *P,V*-plane (pressure-specific volume): *CP* – gas-liquid critical point; *Triangles* – ab initio calculation of crystal phase at $T = 0$ [12] (Figure from [11]).

**FIGURE 4.** "Cold curve" ($U(V)$ at $T = 0$) and critical isotherm ($PV$ at $T = T_{cr}$) vs. $r_S \sim V^{1/3}$ in the modified OCP model (OCP(~)). Notations: *Solid circle* – critical point; *Open circle* – 'normal' state ($T = 0$, $P = 0$); *Solid triangle* – solid spinodal $(\partial P/\partial V)_T = 0\}$; $r_S$ –Bruckner parameter of background electrons (Figure after [7])

The same scenario is incorporated in construction of some semi-empirical equation of state (EOS) of metals [11]. Hypothetical multi-phase density-pressure diagram of Al exposed at Figure 2 as an example of such EOS. Fundamental feature of this scenario of zero-temperature metastable melting is a location of crystal-fluid phase transition with a finite density gap just on the thermodynamically stable repulsive part of "cold curve" i.e. isotherm $T = 0$. Such hypothetical 'cold curve' for Al [11] is shown on Figure 3. It should be stressed that isotherm $T = 0$ coincides with the isentrope $S = 0$, therefore the exposed 'cold' melting at $T = 0$ [11] occurs without entropy change.

# Melting in Standard One Component Plasma Model OCP(#)

Standard variant of one component plasma model on rigid background (following notation – OCP(#)) is a system of ions on rigid compensative background. It contains only one phase transition – Wigner crystallization, which takes place without any density gap i.e. melting zone is just a one-dimensional curve. Melting line of crystal and freezing line of fluid coincides with each other. There are two well-known regime of melting in the model OCP(#):

1) Classical melting of non-degenerated ions in low density limit ($\theta \equiv kT/\varepsilon_F \gg 1$) In this case thermodynamic properties of OCP(#) depend only on dimensionless 'non-ideality' parameter $\Gamma \equiv (4\pi n/3)^{1/3}(Ze)^2/kT$. Melting corresponds to: $\Gamma = \Gamma_{melt} \approx 175$

2) "Cold" quantum melting of highly degenerated ions in high density limit [13] ($\theta \equiv kT/\varepsilon_F \gg 1$). In this case thermodynamic properties of OCP(#) depend on Bruckner 'non-ideality' parameter $r_S \equiv (3/4\pi na)^{1/3}$ Melting corresponds to: $r_S \approx const \approx 100$ [14,15].

General diagram of melting in OCP(#) is shown at Figure 5. It should be stressed that melting in OCP(#) takes place at any temperature and formally follows Simon's law (1) with ($A = const., C = 4$ and $P^* \; 0$).

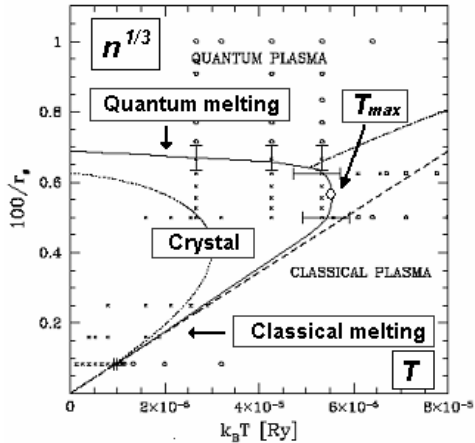 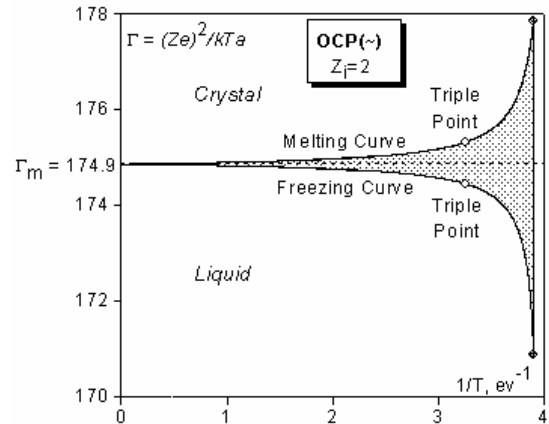

**FIGURE 5.** Boundary of Wigner crystal at OCP(#) with 'classical' and 'quantum' melting and hypothetical pseudocritical point of maximal melting temperature $T_{max}$. *Solid* – melting after [14]; *dash-dotted line* – melting line after [15]; *dashed line* – classical melting limit $\Gamma_{melt} \approx 175$ (Figure from [14]).
**FIGURE 6.** Melting zone of Wigner crystal in modified OCP(~) model [6] with splitted off melting and freezing curves and finite density gap, and low-density melting 'spinodal catastrophe' (From [7])

# Scenarios of Metastable Melting in Modified OCP Model with Uniformly Compressible Background – OCP(~).

More realistic than OCP(#) is a model of one-component plasma with *uniform*, but *compressible* background. This property of the background can be defined self-consistently [16] and can be achieved, for example, with the use of *uniform* ideal Fermi-gas of electrons as a background for OCP positive ions. Due to *uniform compressibility* of background not one, but three 1st-order phase transitions appears in this model namely melting, evaporation and sublimation [6]. Melting in OCP(~) became a two-dimensional zone bounded by melting and freezing lines with finite density gap (Figure 6). Modified OCP model (OCP(~)) introduces principally new behavior of melting 'strip' in the limit $T \to 0$ i.e. in all the cases (see below) the melting zone *does not reach* the $T = 0$ isotherm (!). As a result the 'cold curve' in OCP(~) is smooth and continuous without any phase transition (Figure 4). As for the closure of metastable melting, there may be three variants of its termination in OCP(~) depending on charge number of ions, $Z$. All the variants occur at finite temperature. In first two variants (A) and (B), metastable melting terminates at the liquid or crystal spinodals. In third variant (C) melting transmitted continuously into sublimation, so that there is no metastable melting at all.

The formal definition of Helmholtz free energy in the OCP(~) model is [6]:

$$\left(\frac{F}{NkT}\right)_{OCP(\sim)} \equiv f = \frac{F^{id(e)gas}}{N_e kT} + \frac{\Delta F^{OCP(\#)}}{NkT} = f^{ig(e)gas} + \Delta f^{OCP(\#)} \quad (2)$$

The advantage of the OCP(~) model is due to prohibition (by definition) of any individual ion-electron correlation features of all phase transitions including characteristic of metastable branches can be directly calculated if properties of two individual subsystems namely ions with rigid background i.e. OCP(#) of ions and that of ideal Fermi-gas of electrons are known. Currently both the constituents are well studied via Monte Carlo and Molecular Dynamics simulations [5,14,15] etc. and effective analytical fits are exist, so that all the properties in 'non-associative' OCP(~) models could be calculated explicitly [6, 7, 9, 17].

## Standard Scenario of Metastable Melting Termination in OCP(~)

Along with conventional scenario of metastable melting exposed at Figures 1-3 there is a fundamentally different ones which occur in the OCP(~) model. The first one (**A**) corresponds to the low values of ionic charge number $Z << Z^* \approx 35$ (see [18]) According to this variant in the limit of $T \to 0$ the boundary of freezing of metastable liquid touches a liquid spinodal at finite temperature. While the melting zone is approaching a boundary of liquid state absolute instability, i.e. liquid spinodal, its width is dramatically increasing. Figure 7 and 8 show brief and close view of this event i.e. crossing the freezing curve of metastable fluid and fluid spinodal. At all temperatures below the temperature of this intersection fluid is absolutely thermodynamically instable and doesn't exist even in metastable state. Thus the possibility of melting as physical phenomenon vanishes. The only possibility in this case is spontaneous decomposition of metastable system to thermodynamically stable two-phase mixture of solid and vapour at *finite temperature and pressure*!

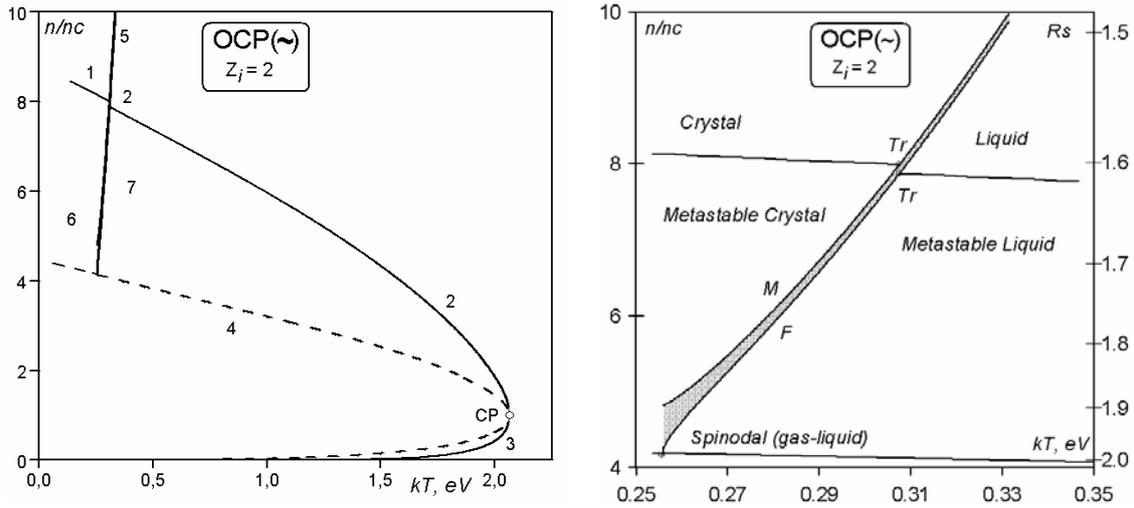

**FIGURE 7.** Spinodal decomposition of metastable melting in the limit of zero temperature. <u>Notation</u>: *1* – crystal; *2* – Liquid; *3* – Gas; *4* – Spinodal $\{(\partial P/\partial V)_T=0\}$; *5* – melting zone; *6* - metastable crystal; *7* – metastable liquid; CP – critical point (Figure from [7])

**FIGURE 8.** The same as Figure 6 in details. <u>Notation</u>: *M* – melting curve; *F* – freezing curve; *Tr* – triple point; *Spinodal(g-l)* – liquid spinodal $\{(\partial P/\partial V)_T=0\}$ (Figure from [7]).

## Anomalous Scenarios of Metastable Melting Termination in OCP(~)

One of the methodical advantages of the OCP(~) model is additional model parameter – charge of ions Z. While the ion charge is increasing the topology of phase diagram is dramatically changing. At the small value of Z the model shows traditional view of phase diagram (see Figure 1 in [18]). But as the ionic charge number *Z* increases the melting zone moves toward the critical point of the fluid-crystal phase transition, passes it and finally shifts on the binodal of low-density phase.

In these cases there are additional two completely new scenarios (B) and (C) of metastable melting:

**Scenario B**. At large values of charge number *Z* ($Z >> Z^{**} >> 45$) the melting zone crosses a low-density slope of fluid-crystal co-existence curve (see Figure 1 in [18]). In this cases, as well as in the case (**A**) described above, the metastable melting zone also doesn't reach the zero isotherm (*T*=0) due to intersection of *melting line* of metastable crystal with *vapor spinodal* (line $\Gamma \approx 6$ at Figure 1 [18]).

**Scenario C**. This case corresponds to intermediate values of ionic charge number $35 \approx Z^* > Z > Z^{**} \approx 45$ (see [18]). In this case the melting line $\Gamma \approx 175$ of model-prototype OCP(#) falls just into the region of critical point of new 1$^{st}$ order phase transition of gas-liquid type in OCP(~). The model shows an exotic 'unified' boundary of unique crystal-fluid phase coexistence, which is continuous superposition of normal melting and sublimation equilibrium (crystal-fluid and crystal-gas) with common boundary of crystal-fluid two-phase area when melting smoothly turns to sublimation (see Figure 4 in [18]). In context of presently discussing problem of possible variants of metastable melting termination it means that the metastable melting as a phenomenon is absent in the system.

# CONCLUSIONS

It should be emphasized that all the details of discussed above hypothetical scenarios of termination of metastable melting in the limit $T \to 0$ could be directly examined in numerical Monte Carlo and Molecular Dynamics simulations. The main problem, which should be overcome on this way, is the necessity of creating and preserving during a finite period of time a deep metastable state for both the competing phases, solid and liquid, with a deep negative pressure in the investigated simulation cell. Experience currently accumulated in such numerical simulation of simultaneous two-phase coexistence (see for example [19]) allows for hope that quick progress in this problem will be achieved soon.

# ACKNOWLEDGEMENTS

The authors are grateful to Hugh DeWitt, G. Kanel, D. Yakovlev and A. Potekhin for helpful discussions.